\documentclass{article}

\usepackage{microtype}
\usepackage{graphicx}
\usepackage{subfigure}
\usepackage{booktabs} % for professional tables

\usepackage{hyperref}
\usepackage{aas_macros}

\usepackage[accepted]{icml2023}

\usepackage{amsmath}
\usepackage{amssymb}
\usepackage{mathtools}
\usepackage{amsthm}

\usepackage[capitalize,noabbrev]{cleveref}

\theoremstyle{plain}

\theoremstyle{definition}

\theoremstyle{remark}

\usepackage[textsize=tiny]{todonotes}

\icmltitlerunning{Multi-input CNN for bright transient identification}

\begin{document}

\twocolumn[
\icmltitle{$\texttt{BTSbot}$: A Multi-input Convolutional Neural Network to Automate and Expedite Bright Transient Identification for the Zwicky Transient Facility}




\begin{icmlauthorlist}
\icmlauthor{Nabeel Rehemtulla}{NU,CIERA}
\icmlauthor{Adam A. Miller}{NU,CIERA}
\icmlauthor{Michael W. Coughlin}{UMinn}
\icmlauthor{Theophile Jegou du Laz}{Caltech}
\icmlauthor{on behalf of ZTF}{}
\end{icmlauthorlist}

% Nabeel ORCID: 0000-0002-5683-2389
% Adam ORCID: 0000-0001-9515-478X
% Michael ORCID: 0000-0002-8262-2924
% Theo ORCID: 0009-0003-6181-4526

\icmlaffiliation{NU}{Department of Physics and Astronomy, Northwestern University, 2145 Sheridan Road, Evanston, IL 60208, USA}
\icmlaffiliation{CIERA}{Center for Interdisciplinary Exploration and Research in Astrophysics (CIERA), 1800 Sherman Ave., Evanston, IL 60201, USA}
\icmlaffiliation{UMinn}{School of Physics and Astronomy, University of Minnesota,
Minneapolis, Minnesota 55455, USA}
\icmlaffiliation{Caltech}{Division of Physics, Mathematics, and Astronomy, California
Institute of Technology, Pasadena, CA 91125, USA}

\icmlcorrespondingauthor{Nabeel Rehemtulla}{nabeelr@u.northwestern.edu}

\icmlkeywords{Machine Learning, ICML}

\vskip 0.3in
]



\printAffiliationsAndNotice{} % otherwise use the standard text.

\begin{abstract}
The Bright Transient Survey (BTS) relies on visual inspection (``scanning") to select sources for accomplishing its mission of spectroscopically classifying all bright extragalactic transients found by the Zwicky Transient Facility (ZTF). We present $\texttt{BTSbot}$, a multi-input convolutional neural network, which provides a bright transient score to individual ZTF detections using their image data and 14 extracted features. $\texttt{BTSbot}$ eliminates the need for scanning by automatically identifying and requesting follow-up observations of new bright ($m\,<18.5\,\mathrm{mag}$) transient candidates. \texttt{BTSbot} outperforms BTS scanners in terms of completeness (99\% vs. 95\%) and identification speed (on average, 7.4~hours quicker).

\textbf{See \textcolor{blue}{\href{https://iopscience.iop.org/article/10.3847/1538-4357/ad5666}{here}} for the full \texttt{BTSbot} publication}
\vspace{-10pt}

\end{abstract}

\section{Introduction}
\label{sec:intro}

Large, wide-field surveys have recently revolutionized time-domain astronomy by repeatedly imaging the entire night sky. These surveys produce staggering amounts of data every night, an influx that demands the adoption of machine learning (ML) techniques. ML models have been applied to a variety of tasks in astronomy including real/bogus classification \citep[e.g., ][]{Bloom+2012, Brink+2013}, photometric transient classification \citep[e.g., ][]{Villar+2020}, photometric redshift estimation \citep[e.g.,  ][]{CarrascoKind+2013}, and many others. For the most part, ML models in astronomy perform their task using a small number of extracted numeric features. Limiting these models to extracted features ignores potentially valuable information present in the associated images. A comparatively small number of convolutional neural networks (CNNs) have been built that make use of the information embedded in astronomical images, and they have generally had great success \citep[e.g., ][]{Dieleman+2015, Lanusse+2018, Duev+2019}. CNNs are particularly well suited to astronomy because they can capture properties, like galaxy morphology, which often remain largely obscured to other image processing techniques. Only a very small subset of these CNNs are \textit{multi-input}, meaning they take in images and input of another type, like extracted features \citep{Carrasco-Davis+2021}. Multi-input CNNs (MI-CNNs) have more and varied information to draw from compared to analogous single-input models. Here, we present a new MI-CNN for bright transient identification.

The Bright Transient Survey \citep[BTS;][]{Fremling+2020, Perley+2020} aims to classify all extragalactic transients with $m < 18.5\,\mathrm{ mag}$ in $g$ or $r$ band from the public alert stream of the Zwicky Transient Facility \citep[ZTF;][]{Bellm+2019}. Every night, experts inspect candidate BTS sources, a process called ``scanning,'' and select the bright extragalactic transients\footnote{We use ``supernova" interchangeably with ``extragalactic transient." These are not equivalent, but it simplifies the prose.} for spectroscopic observation and classification. There are typically $\sim$50 BTS candidates per night, of which $\sim$10 are real bright supernovae (SNe), with the others being mostly dim SNe, active galactic nuclei (AGN), and cataclysmic variables (CVs). Since its origin, BTS has maintained near-perfect completeness of relevant sources and has rapidly and publicly released their classifications: a monumental service to the community. The BTS sample enables an enormous amount of science including some of the largest SN population studies ever conducted. The MI-CNN introduced here, dubbed $\texttt{BTSbot}$, is built to automate scanning for BTS by performing binary classification (bright SN / not bright SN) on ZTF alert packets.

\section{Model scope, architecture, and training}
\label{sec:model-intro}

\begin{figure*}[ht]
% \begin{center}
\centerline{\includegraphics[width=1.7\columnwidth]{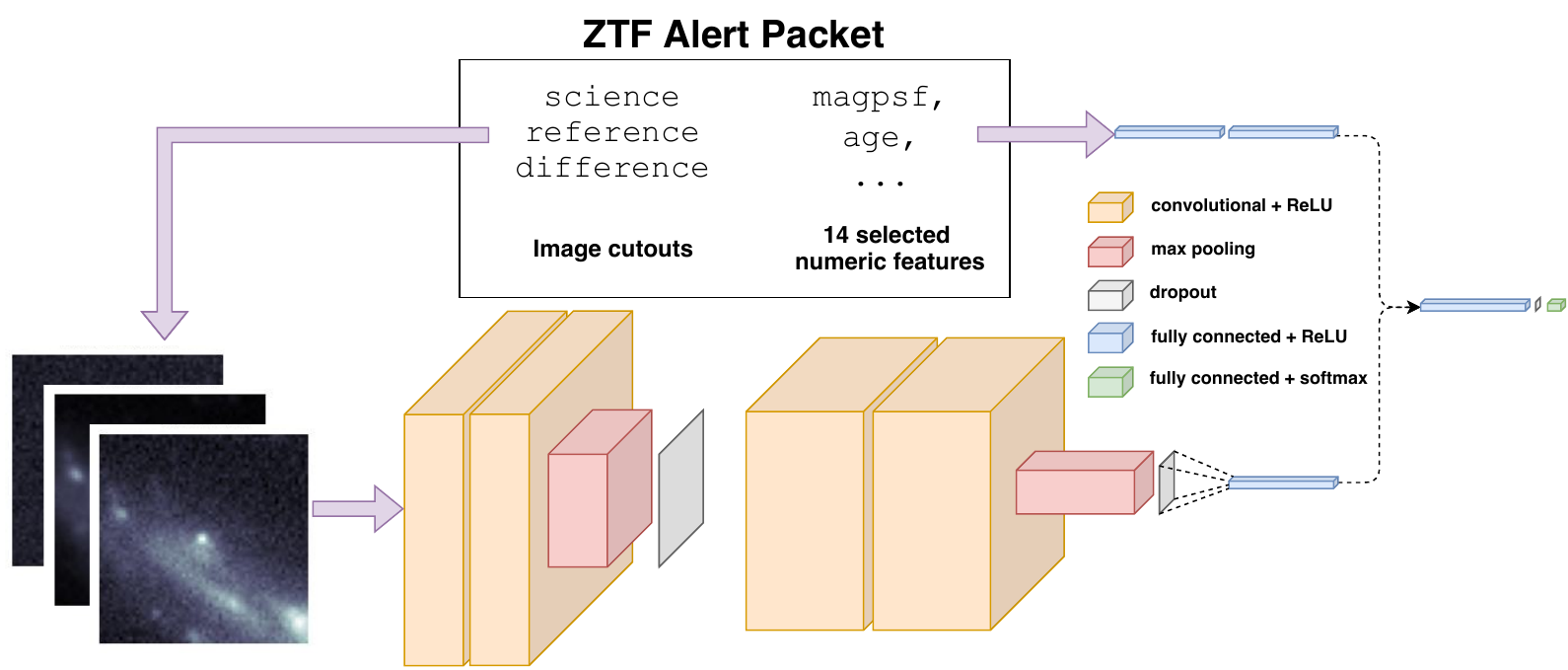}}
\caption{Diagram of \texttt{BTSbot}, a multi-input convolutional neural network, showing the layers that the input images and extracted features pass through before being rejoined and producing the score for bright extragalactic transient binary classification.}
\label{fig:model_diagram}
% \end{center}
\end{figure*}

Our MI-CNN, $\texttt{BTSbot}$, automates BTS scanning by assigning each individual ZTF alert packet a bright transient score. Alerts from real bright SNe will have high scores, allowing us to automatically select them from the pool of BTS candidates for follow-up and classification. Further, our model identifies sources and requests follow-up before a human can, helping to maintain the survey's near-perfect completeness by obtaining a spectrum before the transient fades beyond detection.

Figure~\ref{fig:model_diagram} shows how input is fed into $\texttt{BTSbot}$ and how the information from the extracted features and images is combined to produce the output. $\texttt{BTSbot}$ contains three main components: (1) The convolutional branch processes the science, reference, and difference images as a three-channel image through a VGG-like architecture \citep{Simonyan+2014} followed by flattening. (2) The fully-connected branch processes the extracted features through two dense layers. (3) The combined section concatenates the output of the two branches and passes it through two more fully-connected layers; the second of which produces the final output using a softmax activation function. The output is a unit-interval score where higher scores represent increased confidence that the source in the input alert packet is, or will become, a bright extragalactic transient.

The choice of a MI-CNN is motivated by the fact that the images and the extracted features provide different, valuable information for performing our task. For example, the features \texttt{distpsnr1} and \texttt{sgscore1} respectively represent the angular distance to and the star-galaxy score \citep{Tachibana_Miller_2019} of the Pan-STARRS1 \citep[PS1;][]{Kaiser+2002} cataloged source nearest to this ZTF source. While the new SNe are not present in PS1, their host galaxies often are. Thus, alerts from bright SNe tend to have moderate \texttt{distpsnr1} and small \texttt{sgscore1} values, indicating a galaxy projected nearby to the source. In contrast, AGN and CVs will typically be cataloged in PS1 and thus have \texttt{distpsnr1} very near to zero. The images also provide important information following a similar heuristic. Bright SNe tend to be associated with prominent off-center extended sources (their host galaxies); faint SNe tend to have less prominent host galaxies because they tend to be farther away; AGN will appear as exactly centered extended sources; CVs will often appear surrounded by many bright point sources because they tend to occur near dense star fields. A MI-CNN is able to pool information from all input types and consider them together when making a prediction. Where a single-input CNN might fail due to a lack of discriminating information, a MI-CNN leverages additional and distinct information when making its prediction. For example, a very faint alert that shows next to no features in its images can be identified as an AGN by a MI-CNN because AGN often have a large number of previous detections at their location (represented by \texttt{ndethist}), information a single-input CNN could not have.

Given the scope and architecture of $\texttt{BTSbot}$ we encounter a number of challenges. First, we are requiring the model to learn multiple non-trivial separations. $\texttt{BTSbot}$ must learn to separate SNe from other sources without using distance information because it is not known \textit{a priori}. It must also learn to identify bright SNe from a single alert irrespective of the SN's current phase. Early in its rise or late in its fade, a bright SN can appear very similar to a near-peak dim SN. This is related to the second complication, which is that $\texttt{BTSbot}$ uses no time-series information. Although light curves contain crucial information for evaluating a source as a bright SN or not, we make the choice to omit time-series information, in part, because it would introduce complexity and, likely, noise. There is no established method for representing partial light curves of the wide variety our model encounters in a way that is fit for input into a neural network. There has been a great deal of work to accomplish this for SNe alone \citep{Villar+2020}, but these methods are not applicable to all the types of sources that our model encounters. Further, omitting time-series information offers \texttt{BTSbot} an advantage over light curve-based models. With its alert-based architecture, \texttt{BTSbot} can identify a bright SN from its very first detection, rather than preferring multiple detections to begin constraining a light curve. We expect that a similar model with light curve information would identify bright SNe as or less quickly than $\texttt{BTSbot}$ does, potentially hurting BTS's completeness. Section~\ref{sec:2023ixf} illustrates how prompt and automatic identification of SNe with $\texttt{BTSbot}$ can aid in uncovering poorly understood SN physics. To our knowledge, no other CNNs exist that simultaneously filter out non-SNe and predict a SN's future behavior given a single snapshot in time.

% \begin{figure}[]
% % \begin{center}
% \centerline{\includegraphics[width=0.85\columnwidth]{figs/score_dist.pdf}}
% \caption{2D-histogram of the model's score versus the input alert's PSF magnitude with marginals above and to the left. We observe that the model correctly learns that the score should be magnitude-dependent. We also observe a population of alerts with $m~<~18.5\,\mathrm{mag}$ and $\textrm{score}\approx0$ that break this trend. These are mostly AGN and thus correctly classified. The model learns to select bright extragalactic transients and reject other sources.}
% % \end{center}
% \label{fig:score_dist}
% \end{figure}

\begin{figure}[]
% \begin{center}
\centerline{\includegraphics[width=0.9\columnwidth]{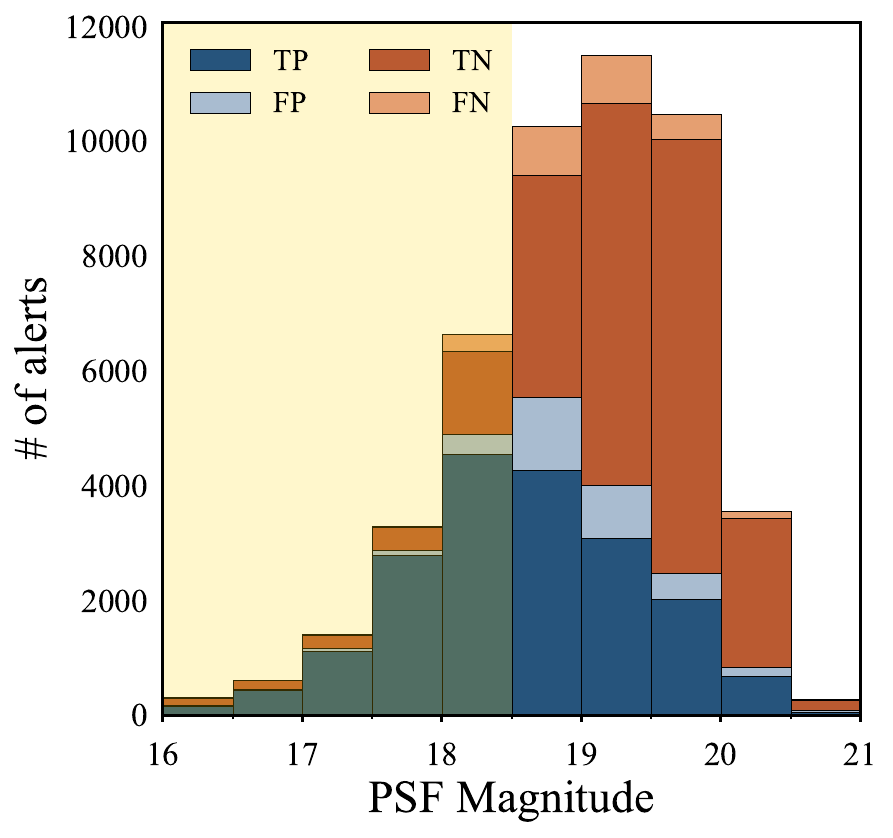}}
\caption{Stacked bar chart showing the distribution of classification outcomes as a function of the input alerts' PSF magnitude. The highlighted region to the left is most important because it marks the alerts which BTS scanners use to trigger spectroscopic observations. Performance of alerts with $m<18.5\,\mathrm{mag}$ is excellent and misclassifications are mostly in the dimmest bin.}
% \end{center}
\label{fig:classification_type}
\end{figure}

\begin{figure*}[ht!]
% \begin{center}
\centerline{
    \includegraphics[width=0.875\columnwidth]{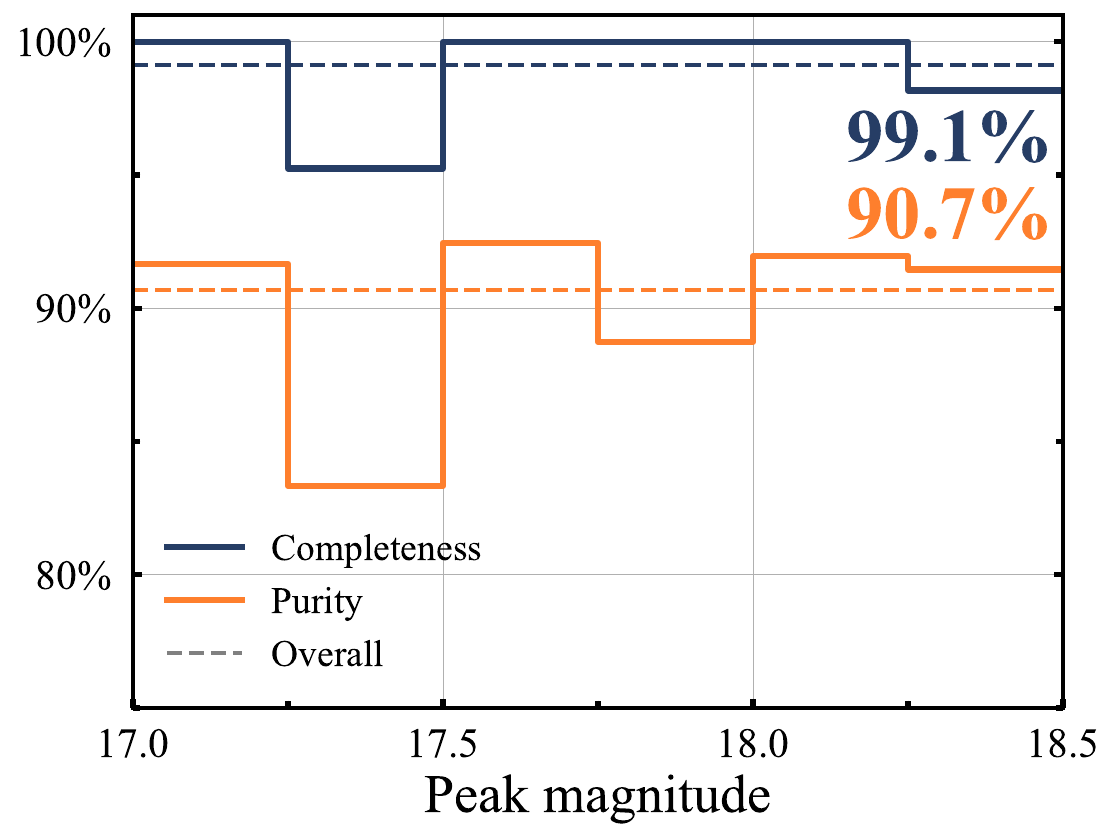}
    \includegraphics[width=0.85\columnwidth]{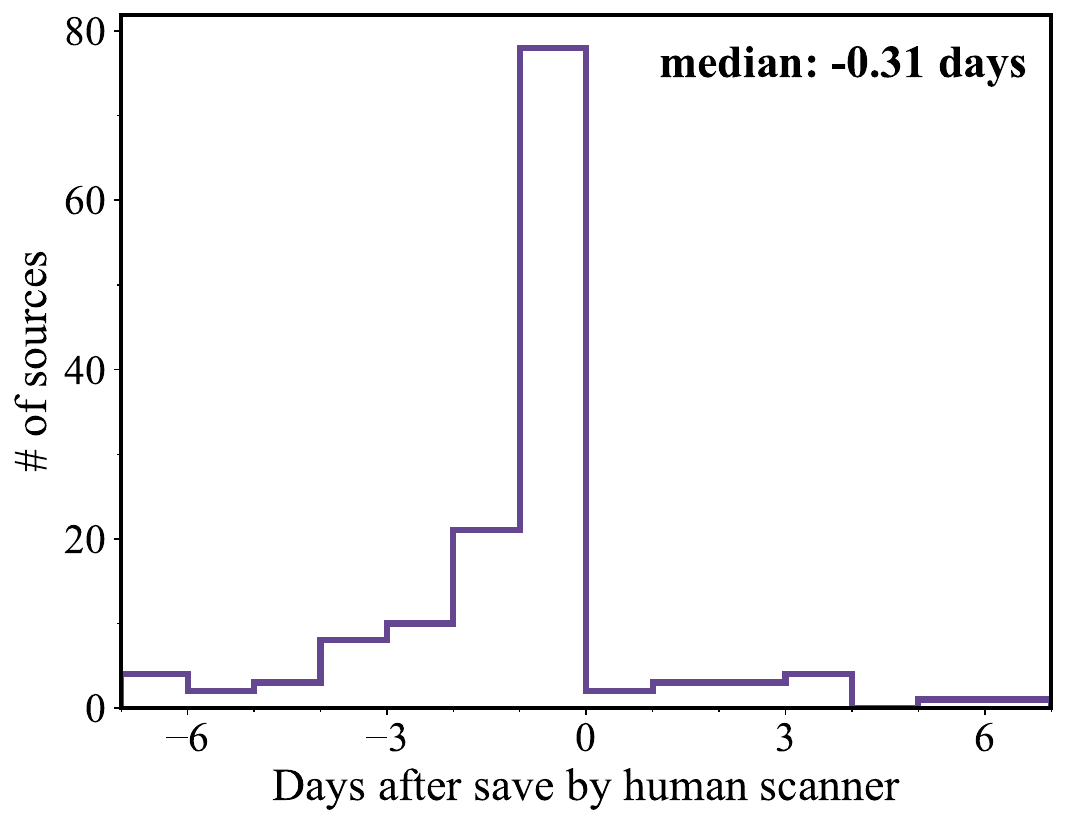}
}
\caption{\texttt{BTSbot} performance when classifying sources as bright supernovae given at least one alert with high score. Left panel: completeness (navy) and purity (orange) as a function of peak magnitude and respective averages (dashed lines). \texttt{BTSbot} produces a sample that is $99.1\%$ complete and $90.7\%$ pure overall. The bright limit is noisy due to the small number of sources in those bins. Right panel: distribution of delay between \texttt{BTSbot} and BTS scanner identification times with median at -0.31 days, indicating that, on average, \texttt{BTSbot} identifies sources about 7 hours quicker than BTS scanners.}
\label{fig:gt1}
% \end{center}
\end{figure*}

\subsection{Training}
\label{sec:training}

Solving this complex classification task requires a significant training set. Having run since 2018, BTS has now amassed the largest set of public SNe classifications ever. The size of this labeled data set enables the construction of \texttt{BTSbot}. Bright SNe classified by BTS populate the positive class, while AGN, CVs, and dim SNe rejected by BTS constitute the negative class. In total, we have 561,000 alerts ($\sim$44\% of which belong to the positive class) from 14,348 sources ($\sim$33\% of which belong to the positive class). This difference in class distribution between alerts and sources stems from some sources having many more alerts than others: AGN can have thousands, bright SNe typically have dozens, and dim SNe can have as little as a few. Training on every alert would result in some types of sources being over-represented. To remedy this we define a hyperparameter called $N_\textrm{max}$, the maximum number of alerts included per source. We find $N_\textrm{max}=60$ to be optimal; it balances between thinning extra alerts from sources like AGN and maximizing the training set size. We also weight contributions to the loss function by the relative size of each class to mitigate the effects of class imbalance.

$\texttt{BTSbot}$ is trained with mostly standard hyperparameters. We adopt the Adam optimizer \citep{Kingma+2014} and the binary cross-entropy loss function. We perform a number of Bayesian hyperparameter searches with the Weights and Biases platform \cite{wandb} to optimize our choice of batch size, Adam parameters, $N_\textrm{max}$, which extracted features to include, and more. We also employ data augmentation to the image cutouts like random rotations of 0$^\circ$, 90$^\circ$, 180$^\circ$, and 270$^\circ$ and random horizontal and vertical flipping. These help prevent overfitting and ensure that $\texttt{BTSbot}$ is invariant to these transformations.

We employ a standard 81/9/10\% train/validation/test split. We also prevent sources from having alerts in multiple splits. This prevents different splits from containing extremely similar data, which would introduce bias.

\section{Model performance}

The trained model is run on the full validation split (i.e. no $N_\textrm{max}$ thinning) to create unbiased performance diagnostics.

% Figure~\ref{fig:score_dist} shows the relationship between the raw scores and the input alerts' PSF magnitudes. The distribution reveals a magnitude-dependence for the scores, demonstrating that the model is learning that brighter alerts tend to be from bright supernovae (SNe). This trend is broken by a collection of alerts with $m<18.5\,\mathrm{mag}$ and scores very near to zero. However, these are not misclassifications; the vast majority of these are correctly classified bright AGN. Together, these demonstrate that the model successfully learns to identify bright SNe while rejecting bright non-SNe and other dim sources. The significance of the hot-spot around $\textrm{score}\approx0.45$ and $m>18.5\,\mathrm{mag}$ is not immediately clear. It may suggest that the model learns to be uncertain for these alerts.  

Figure~\ref{fig:classification_type} illustrates classification outcomes as a function of PSF magnitude. The highlight marks $m < 18.5\,\mathrm{mag}$, the BTS magnitude threshold. The alerts within the highlight are especially important to classify correctly because they influence whether or not a spectrum of the source is to be collected. In that region, misclassifications are very limited and mostly in the dimmest bin. For $m > 18.5\,\mathrm{mag}$, there are more misclassifications and many more alerts overall. 

Ultimately, alert-based metrics are not perfectly representative of the model's real-world performance; source-based classification is more relevant. The chosen metrics must consider that the model has multiple chances to correctly (or incorrectly) classify a source. To this end, we define a ``policy" that maps the real-time history of a source's scores to a source-based classification. For simplicity, we put aside policy optimization for now and only consider one naive policy with the impression that we could improve performance with a more sophisticated choice of policy.

Our naive policy is named \texttt{gt1}; it classifies a source as a bright SN once it has \textit{one} alert with score $\geq0.5$. The left panel of Figure~\ref{fig:gt1} shows the purity and completeness of the sample produced when following this policy. We observe that \texttt{gt1} yields a sample with $99.1\%$ completeness and $90.7\%$ purity overall. BTS scanners produce a sample that is roughly $95\%$ in both completeness and purity \cite{Perley+2020}. In tests with other more conservative policies like \texttt{gt3} (defined analogously to \texttt{gt1}), we observe that purity is increased at the cost of completeness. We favor \texttt{gt1} because it maximizes completeness, BTS's highest priority. The right panel of Figure~\ref{fig:gt1} compares the time between when \texttt{BTSbot} and the BTS scanners identified some source. Here, negative numbers represent that \texttt{BTSbot} identified the bright SN before the scanners and positive numbers represent the opposite. The median is -0.31 days, indicating that \texttt{BTSbot} expedites the identification of bright SNe by, on average, 7.44 hours. Expediting identification by several hours will frequently mean observing the source an entire night earlier, thus yielding an improvement of $\sim$24 hours in practice. With \texttt{gt1}, \texttt{BTSbot} outperforms human scanners in completeness and speed, while making only a small compromise with slightly lower purity.

\section{Real-time, real-world usage}
\label{sec:usage}

\texttt{BTSbot} has been integrated into Fritz, the first-party ZTF alert-broker and a SkyPortal instance \cite{Coughlin_2023}, and is currently posting scores to all new ZTF alert packets. Alert packets are created and augmented with a bright transient scores just minutes after the observations are taken with ZTF. During the night, a tool we call \texttt{autoscan} checks for new BTS candidates every half hour and immediately saves those that pass \texttt{gt1}. \texttt{autoscan} will soon also simultaneously request follow-up for passing sources. With this, we can monitor the model's real-time, real-world performance and note any frequent misclassifications. We have been awarded a large observing program for the 2023B semester with the SEDMv2 spectrograph on the Kitt Peak 2.1 m telescope. With this time, we will build a spectroscopic sample that represents the model's unbiased performance. The resulting performance metrics will be invaluable when considering adaptations of this MI-CNN, e.g., to specific SN sub-types or exotic transients like kilonovae.

\subsection{SN~2023ixf \& Rapid follow-up with \texttt{BTSbot}}
\label{sec:2023ixf}

We take SN\,2023ixf, the recent Type-II supernova in M101, as an example illustrative of the additional discovery potential of an alert-based model like \texttt{BTSbot}.
%Although its first-light was even earlier \cite{Mao_ixf_TNS}, 
SN\,2023ixf was reported to TNS by Koichi Itagaki at 14:42\,PDT on May 19th \cite{Itagaki_ixf_TNS}.
%, and its time of first-light was later constrained to be about 25 hours earlier \cite{Mao_ixf_TNS}. 
The earliest published spectrum was collected by Daniel Perley less than an hour later with the SPRAT spectrograph \cite{Perley_ixf_TNS}. About 14 hours before the first TNS report, SN\,2023ixf was detected by ZTF, and, just minutes later, this alert packet was assigned a bright transient score of 0.840 by \texttt{BTSbot}. If it was in-place at the time, \texttt{autoscan} would have identified this new source passing \texttt{gt1} at about 01:00\,PDT and requested a spectrum from one of the numerous robotic spectrographs associated with ZTF and BTS, e.g., SEDM. At this point, there is about half of the night remaining at SEDM's location, plenty of time for a spectrum to be collected. Even if the spectrum is collected at the end of the night ($\sim$05:00\,PDT), this represents a $\sim$10 hour speed-up over the otherwise earliest spectrum. In this example, \texttt{BTSbot} and \texttt{autoscan} probe the early, rapidly-evolving explosion physics mostly unavailable to traditional triggering methods. This would not be possible with light curve-based models, which typically require using the source's evolution over multiple days to identify it as a transient. This extremely rapid follow-up is enabled by the alert-based architecture of \texttt{BTSbot}.

\section*{Software and Data}
Code relating to \texttt{BTSbot} and \texttt{autoscan} are made public at \url{https://github.com/nabeelre/BTSbot}. In their development, we use Astropy \citep{astropy:2013, astropy:2018}, cron \cite{cron}, Keras \cite{keras}, Matplotlib \citep{hunter07}, NumPy \citep{numpy}, pandas \cite{pandas1, pandas2}, penquins \cite{penquins}, scikit-learn \cite{sklearn}, and Tensorflow \cite{tensorflow}.

\section*{Acknowledgements}

The material contained in this document is based upon work supported by a National Aeronautics and Space Administration (NASA) grant or cooperative agreement. Any opinions, findings, conclusions, or recommendations expressed in this material are those of the author and do not necessarily reflect the views of NASA. This work was supported through a NASA grant awarded to the Illinois/NASA Space Grant Consortium.

M.~W.~Coughlin acknowledges support from the National Science Foundation with grant numbers PHY-2010970 and OAC-2117997.

Based on observations obtained with the Samuel Oschin Telescope 48-inch and the 60-inch Telescope at the Palomar Observatory as part of the Zwicky Transient Facility project. Major funding has been provided by the U.S National Science Foundation under Grant No. AST-1440341 and by the ZTF partner institutions: the California Institute of Technology, the Oskar Klein Centre, the Weizmann Institute of Science, the University of Maryland, the University of Washington, Deutsches Elektronen-Synchrotron, the University of Wisconsin-Milwaukee, and the TANGO Program of the University System of Taiwan.

\bibliography{paper}
\bibliographystyle{icml2023}

%%%%%%%%%%%%%%%%%%%%%%%%%%%%%%%%%%%%%%%%%%%%%%%%%%%%%%%%%%%%%%%%%%%%%%%%%%%%%%%
%%%%%%%%%%%%%%%%%%%%%%%%%%%%%%%%%%%%%%%%%%%%%%%%%%%%%%%%%%%%%%%%%%%%%%%%%%%%%%%
% APPENDIX
%%%%%%%%%%%%%%%%%%%%%%%%%%%%%%%%%%%%%%%%%%%%%%%%%%%%%%%%%%%%%%%%%%%%%%%%%%%%%%%
%%%%%%%%%%%%%%%%%%%%%%%%%%%%%%%%%%%%%%%%%%%%%%%%%%%%%%%%%%%%%%%%%%%%%%%%%%%%%%%
% \newpage
% \appendix
% \onecolumn

% You can have as much text here as you want. The main body must be at most $8$ pages long.
% For the final version, one more page can be added.
% If you want, you can use an appendix like this one, even using the one-column format.
%%%%%%%%%%%%%%%%%%%%%%%%%%%%%%%%%%%%%%%%%%%%%%%%%%%%%%%%%%%%%%%%%%%%%%%%%%%%%%%
%%%%%%%%%%%%%%%%%%%%%%%%%%%%%%%%%%%%%%%%%%%%%%%%%%%%%%%%%%%%%%%%%%%%%%%%%%%%%%%

\end{document}